\documentclass[aps,prbbib,twocolumn,epsf]{revtex4}

\usepackage{graphicx}
\usepackage[usenames]{color}
\begin{document}
\draft

\title{Electron beam induced current in photovoltaics with high recombination}
\author{Paul M. Haney$^1$, Heayoung P. Yoon$^{1,2}$, Prakash Koirala$^3$, Robert W. Collins$^3$, Nikolai B. Zhitenev$^1$}

\affiliation{1.  Center for Nanoscale Science and Technology, National Institute of Standards and Technology, Gaithersburg, MD 20899 \\
2.  Maryland NanoCenter, University of Maryland, College Park, MD 20742, USA \\
3.  Department of Physics and Astronomy, University of Toledo, Toledo, OH, 43606, USA}
\begin{abstract}
Electron beam induced current (EBIC) is a powerful characterization technique which offers the high spatial resolution needed to study polycrystalline solar cells.
Ideally, an EBIC measurement reflects the spatially resolved quantum efficiency of the device.  In this work, a model for EBIC measurements is presented which applies when recombination within the depletion region is substantial.  This model is motivated by cross-sectional EBIC experiments on CdS-CdTe photovoltaic cells which show that the maximum efficiency of carrier collection is less than 100 \% and varies throughout the depletion region.  The model can reproduce experimental results only if the mobility-lifetime product $\mu\tau$ is spatially varying within the depletion region.  The reduced collection efficiency is speculated to be related to high-injection effects, and the resulting increased radiative recombination.
\end{abstract}

\maketitle

\section{Introduction}

\begin{figure}[h!]
\begin{center}
\vskip 0.2 cm
\includegraphics[width=9cm]{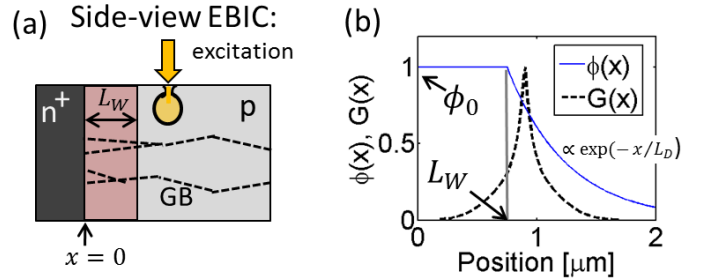}
\vskip 0.2 cm \caption{(a) Schematic of side-view EBIC experiment, and (b) the conventional model for fitting the data.  The generation profile due to the electron beam $G(x)$ is convolved with a collection probability function $\phi(x)$.  The generation profiles depend on position and beam energy.  Larger beam energies result in larger regions of electron-hole pair generation.}\label{fig:1}
\end{center}
\end{figure}

Polycrystalline photovoltaic materials such as CdTe exhibit the remarkable simultaneous properties of high power conversion efficiency and high defect density \cite{kumar}.  Grain boundaries are an important source of defects in these materials.  Quantitative information about the electronic properties at the length scale of individual grains (typically $1~{\rm \mu m}$) is crucial for the further development of these materials.  A measurement technique which offers such spatial resolution is electron beam induced current (EBIC).  Fig. \ref{fig:1}(a) shows a cartoon of an EBIC experiment: electron-hole pairs are created by a beam of high energy electrons in proximity to an exposed surface.  The electrical current is then measured as a function of excitation position \cite{Hanoka}, which determines the carrier collection efficiency.  EBIC has been used as a diagnostic tool for measuring important material properties such as the minority carrier diffusion length and surface recombination velocity \cite{wu,roosbroek,donolato}.  It is generally assumed that all carriers generated in regions with electric fields are collected, as the field rapidly separates electrons and holes before they recombine.  The electric field may be from a Schottky contact, or from the internal field of a {\it p-n} junction (e.g. a depletion region - in Fig. \ref{fig:1}(a), the depletion region is between $x=0$ and $x=L_W$).  Most of the information from EBIC signals is derived from field-free regions, where carriers undergo diffusion and recombination (the region with $x>L_W$ in Fig. \ref{fig:1}(a)).  In these regions, the collection efficiency generally decays exponentially as a function of the distance from the depletion region.  The length scale of this decay is given by the diffusion length $L_D$ \cite{donolato}.  This simple spatial dependence enables an estimate of the diffusion length simply from the {\it lineshape} of the signal - its absolute value doesn't enter into the analysis.  However in this study the absolute value of the EBIC current plays a central role in our analysis.

We next describe EBIC experiments in more detail in order to set the stage for the challenges of EBIC for materials like CdTe.  Fig. \ref{fig:1} (b) shows the model which represents the description of EBIC given above.  The solid blue line $\phi(x)$ is the collection probability for an electron-hole pair generated at position $x$.  The measured current is a convolution of $\phi(x)$ and the excitation profile $G(x)$ (dashed line).  Empirically, the size of the excitation $R_B$ varies with beam energy $E_{\rm beam}$ as $E_{\rm beam}^{1.75}$ \cite{gruen}; for typical beam energies, $R_B$ varies between $100~{\rm nm}$ and $2~{\rm \mu m}$.  Increasing the beam energy results in excitations further away from the surface.  Systematically varying the beam energy enables the separation of surface and bulk contributions to recombination.  Note that the cartoon of the system in Fig. \ref{fig:1}(a) depicts grain boundaries (GB), but the model collection probability is 1-dimensional and assumes a homogeneous material.

The high-resolution capabilities of EBIC are apparently well matched with the requirements of polycrystalline photovoltaic characterization.  However, most studies utilizing EBIC for these materials to date are qualitative in nature.  The development of {\it quantitative} analysis of EBIC in these materials has been hampered due to several factors, some of which we enumerate here: 1. The influence of grain boundaries.  While EBIC has been developed and used to study grain boundaries in Si (determining the recombination velocity for example \cite{palm,donolato2}), the study of grain boundaries in materials like CdTe and Copper Indium Gallium Selenide (CIGS) remains qualitative for the most part.  This is due to the varying influence of grain boundaries in these materials - with proper sample treatment, grain boundaries become passivated and may collect more current than grain interiors \cite{li,modes,yoon}.  Additionally grain boundaries in these materials are charged \cite{fisher}, leading to a complex distribution of electrostatic fields in the device.  A proper account of carrier collection from grain boundaries requires a 3-dimensional deconvolution of the measured current, posing a significant challenge.  2.  Strong surface effects:  to properly resolve grain boundaries at low beam energies, a flat exposed surface is required in order to minimize the effects of roughness on the signal, which can overwhelm the grain boundary contrast.  The most common surface treatment is focused ion beam milling.  This process can introduce significant changes in the surface composition and electronic properties, which can strongly influence the EBIC signal.  The introduction of surface charge by Ga implantation, for example, requires additional modeling beyond a surface recombination velocity boundary condition.  3.  The effect of low mobility and/or low doping combined with the high electron-hole pair generation rate density from the electron beam excitation.  The combination of these two effects may result in a nonlinear response of the system, by for example screening the built-in electric field from the high density of nonequilibrium charges, or leading to an increased contribution from radiative and Auger recombination mechanisms - these processes become important for higher nonequilibrium charge density.

In practice, the EBIC signal contains a convolution of all three of these factors, making it difficult to gain an understanding of any one factor without knowing all three.  We have explored these three factors in detail, and will present the analysis of each in a series of forthcoming papers.  In the current work, we focus on a subset of factor 3 (high injection) by examining the maximum collection efficiency of the EBIC signal for CdTe solar cells.  By carefully estimating the number of electron-hole pairs generated by the electron beam, we find that the maximum value of the collection effiency is well below 1, violating the assumption that all carriers generated within the depletion region are collected.  
A reduced internal quantum efficiency in CdTe has also been observed for optical excitation experiments at strongly absorbed wavelengths \cite{loweff}, where most electron-hole pair generation is within the depletion region.  As we discuss next, any degree of recombination in the depletion region imposes strong constraints on the system and material properties.

For typical material parameters, the charge separation from the internal field $E$ of a {\it p-n} junction is very rapid.  For a built-in field of $10^4~{\rm V/cm}$ and mobility $\mu=10~{\rm cm^2/(V\cdot s)}$, drift velocities of $\mu E=10^5~{\rm cm/s}$ are attained, and free charges separate faster than typical recombination times $\tau$ (typically greater than $1~{\rm ns}$).  Any substantial recombination therefore requires a considerable reduction in $\mu\tau$ and/or in the built-in electric field $E$.  Here we focus on the reduction in collection efficiency due to a reduced value for $\mu\tau$ within the depletion region.  Previous work considers a reduction in $E$ due to screening from nonequilibrium charges in a high charge injection regime \cite{Nichterwitz}.  In this work, we find that recombination in the depletion region is important when the diffusion length $\sqrt{D\tau}$ ($D$ is the diffusivity) is greater than the drift length $\mu E \tau$.  This can also be expressed as $\mu\tau<k_{\rm B}T/qE^2$, where $k_{\rm B}$ is Boltzmann's constant, $T$ is temperature, and $q$ is the absolute value of the electron charge.  For typical material parameters, $k_{\rm B}T/qE^2 \approx 10^{-10}~{\rm cm^2/V}$, so that a strong reduction in $\mu\tau$ is necessary for substantial recombination in the depletion region.

The paper is organized as follows: in Sec. II we present experimental results of EBIC response of CdTe solar cells which demonstrate a maximum collection efficiency of less than 1.  In Sec. III we describe an analytical model of the EBIC response which accounts for recombination within the depletion region, and show that it matches numerical simulations very well.  In Sec. IV, we re-examine the EBIC data with this new model.  We find that in order to match experimental data, $\mu\tau$ must be spatially dependent, reaching a minimum in the depletion region and increasing into the neutral region.  We comment on the possible origin for the reduced efficiency, and consider radiative recombination due to the high generation rate densities associated with the electron beam excitation.

\section{Experiment}\label{sec:expt}

We first describe the cross-sectional EBIC measurements performed on $n^+$ CdS - $p$ CdTe photovoltaic cells.  To indicate the generality of the results, we present data from two rather different samples, with nominal CdTe thicknesses of $1.7~{\rm \mu m}$ and $3~{\rm \mu m}$, and respective power conversion efficiencies of 13 \% and 10 \%.  We refer to the thinner (thicker) sample as ``Sample 1 (2)".  The cross sectional samples are prepared by cleaving.  We have also characterized devices prepared with surface focused ion beam milling (FIB) using Ga ions.  We've found that FIB preparation leads to substantial surface effects which will be described in later work.  As discussed in Appendix A, there is significant lateral variation in the EBIC signal due to the presence of grain boundaries.  We carefully select a linescan from the center of a large grain to minimize the grain boundary influence on the signal.  Acquisition of EBIC at different electron energies was performed with an Indium contact on Indium Tin Oxide (ITO)/$n$-CdS and  a metal probe tip on $p$-CdTe/Cu/Au.  We present results in terms of the EBIC efficiency $\eta$, defined as the ratio of the measured current to the total generation rate of electron-hole pairs $G_{\rm tot}\times q$, where $q$ is the absolute value of the electron charge. (We denote total generation rate by $G_{\rm tot}$, and generation rate density by $G$.)  This total generation rate is estimated as \cite{wu}:
\begin{eqnarray}
G_{\rm tot} = \left(1-b\right)\frac{\left(I_{\rm beam}/q\right)\times \left(E_{\rm beam}/E_0\right)}{2.59 \times \left(E_g/E_0\right) + 0.17}, \label{eq:G}
\end{eqnarray}
 where$I_{\rm beam}$ is the electron beam current, $E_{\rm beam}$ is the beam energy, $E_g$ is the material bandgap, $E_0=1~{\rm eV}$, and $b$ is the backscattering coefficient, corresponding to the fraction of reflected energy \cite{footnote} ($b$ is determined by Monte Carlo calculations).  The beam energy is varied between 5 ${\rm keV}$ and 20 ${\rm keV}$.  The spatial extent of the excitation bulb $R_B$ depends on beam energy through the following empirical relation \cite{gruen}:
\begin{eqnarray}
R_B=\frac{0.043\times R_0}{\left(\rho/\rho_0\right)}\left(E_{\rm beam}/E_0'\right)^{1.75}\label{eq:GR}
\end{eqnarray}
where $\rho$ is the material mass density, $\rho_0=1~{\rm g/cm^3}$, $E_0'=1~{\rm keV}$, and $R_0=1~{\rm \mu m}$.

We estimate 10~\% uncertainty in the measured EBIC efficiency $\eta$ (all uncertainties are reported as one standard deviation).  The dominant sources of uncertainty are from the beam current, and from the inhomogeneous material composition, which introduces uncertainty into the the backscattering coefficient (performed for pure CdTe).  We omit the error bars in the plotted data for clarity, but have included them in the fitting parameters' values.

\begin{figure}[h!]
\begin{center}
\vskip 0.2 cm
\includegraphics[width=9cm]{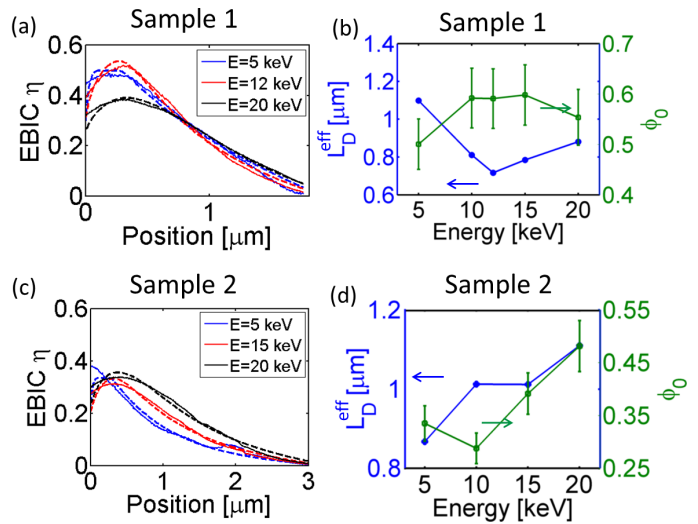}
\vskip 0.2 cm \caption{(a) Solid lines are experimental EBIC profiles for a cleaved CdTe sample.  Dotted lines are model fits.  (b) Shows the energy-dependent fit parameters $L_D^{\rm eff}$ and $\phi_0$. (c) and (d) show the same information for sample 2.}\label{fig:fit1}
\end{center}
\end{figure}

Fig. \ref{fig:fit1}(a) and (c) show the measured EBIC efficiency of CdTe solar cells as a function of distance from the CdS-CdTe metallurgical junction for the two samples.  The profile is taken from a single grain, see Appendix A for full maps of the EBIC response.  The maximum collection efficiency is clearly less than 1 for both samples, and varies throughout the depletion region.  Previous studies of CdTe \cite{poplawsky,durose} and CIGS \cite{Nichterwitz} have also observed a maximum quantum efficiency of below 1, with values similar to, or smaller than those reported here.

To estimate the maximum collection efficiency and diffusion length, we first fit the data to the model depicted in Fig. \ref{fig:1}(b).  This model is well-established and has been applied in many previous studies \cite{donolato,rau}.  Briefly, the model assumes a constant collection efficiency $\phi_0$ within the depletion width $L_W$. $\phi_0$ is usually assumed to be 1, as described in the introduction; here we take $\phi_0$ as a free parameter.  The collection probability decays exponentially from the depletion region edge.  The length scale of this decay is the {\it effective} minority carrier diffusion length $L_D^{\rm eff}$ - which includes recombination contributions from the exposed surface.  The model accounts for recombination from the bulk, the exposed surface, and the back contact \cite{phi_footnote}.  The measured EBIC signal for an electron beam positioned at $x_0$ is the convolution of the collection probability function $\phi(x)$ and the generation rate density profile of electron-hole pairs $G(x,x_0,E_{\rm beam})$ (note that this profile depends on the beam energy $E_{\rm beam}$).  We use a parameterized form for $G(x,x_0,E_{\rm beam})$ from Ref. \cite{rechid}, and have checked that it agrees well with Monte Carlo simulations.  Fig. \ref{fig:1}(b) shows $G(x,x_0,E_{\rm beam})$ for $x_0=0.95~{\rm \mu m}$ and $E_{\rm beam}=15~{\rm keV}$.

We perform least-squares fitting of the data of Fig. \ref{fig:fit1}(a) and (c) with the convolution of $\phi(x)$ and $G(x,x_0,E_{\rm beam})$ to determine $L_W$, $L_D^{\rm eff}$, $\phi_0$, and the back contact recombination velocity $S_C$.
\begin{eqnarray}
\eta\left(x_0\right) = \int \phi\left(x;L_D^{\rm eff},L_W,\phi_0,S_C\right)\times G\left(x,x_0,E_{\rm beam}\right)dx
\end{eqnarray}

Note that $L_D^{\rm eff}$ and $\phi_0$ depend on beam energy because the excitation profile's proximity to the surface (and its associated increased recombination) is energy-dependent.  The variation of $L_D^{\rm eff}$ with the beam energy allows for the extraction of the surface recombination and bulk diffusion length \cite{donolato2,rau}; however we do not present that analysis here.  For both samples we find a depletion width $L_W=\left(0.3\pm0.03\right)~{\rm \mu m}$, which is lower than the depletion width measured with impedance spectroscopy \cite{hamadani}, which ranges from $0.6~{\rm \mu m}$ to $1.2~{\rm \mu m}$, and lower than the expected value given the nominal sample doping ($10^{15}~{\rm cm^{-3}}$).
The maximum collection efficiency increases for high beam energies, however remains well below 1.  At the highest beam energies, surface effects are minimized (the excitation bulb is $1.3 ~{\rm \mu m}$ for $E_{\rm beam}=20~{\rm keV}$).  We therefore conclude that the low efficiency is not primarily due to surface recombination.  In the remainder of the paper, we focus mostly on the maximum EBIC efficiency and the shape of the EBIC response within the depletion region.

As a control experiment, we have performed cross-section EBIC measurements on Si solar cells, and found $\phi_0=1$ within uncertainty, and values of $L_W$ and $L_D$ which agree with the expected results.  As discussed in the introduction, the reduced maximum collection efficiency in CdTe indicates that important physics is missing from the model presented in this section.  In the next section, we present a model which includes recombination in the depletion region.  The data will be re-analyzed in the context of this new model, enabling a more informed analysis of the possible physics responsible for the reduced collection efficiency.

\section{Junction recombination model}\label{sec:model}

\begin{figure}[h!]
\begin{center}
\vskip 0.2 cm
\includegraphics[width=8cm]{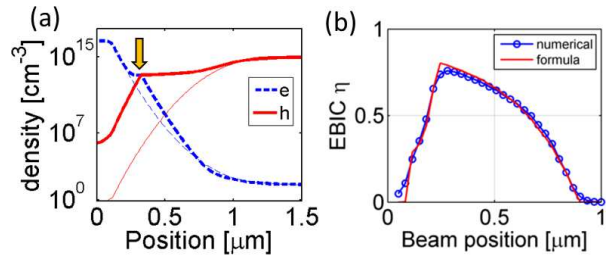}
\vskip 0.2 cm \caption{(a) shows the density of electrons (blue dashed) and holes (solid red). The thinner lines are the equilibrium densities, while the thicker lines are the total densities upon excitation.  The model parameters are: doping density of p-type material is $N_A=10^{15}~{\rm cm^{-3}},$ majority and minority carrier mobilities are $~\mu=10~{\rm cm^2/(V\cdot s)}$, bulk lifetime is $\tau_{\rm bulk}=50~{\rm ns},$ bandgap is $E_g=1~{\rm eV},$ and dielectric constant $\epsilon=11\epsilon_0$.  The junction is located at $x=0.12~{\rm \mu m}$, and the total system length is $3~{\rm \mu m}$.  (b) A comparison between the full simulation results and the analytic formula (Eq. \ref{eq:ebic1}) for the EBIC efficiency. }\label{fig:model1}
\end{center}
\end{figure}

To model the material response to a localized excitation within the depletion region, we consider charge transport arising from a point source excitation in the presence of an electric field $E$.  Fig. \ref{fig:model1}(a) shows the numerical results of a 1-d simulation of a $p$-$n$ junction with an excitation localized at $x_0=0.35~{\rm \mu m}$.  Notice that at the excitation point, the electron and hole densities are equal.  Since the equilibrium carrier concentration in the depletion region is very small, a moderate excitation rate positioned within the depletion region at $x_0$ is sufficient to ensure that $n(x_0)=p(x_0)$.  The charge collection efficiency $\eta$ presented here assumes this condition.  We also assume a uniform electric field $E$.  We defer the detailed derivation to Appendix B, and here present the final result for the charge collection efficiency as a function of the electric field:
\begin{eqnarray}
\eta\left(E\right) =
f\left(E\right)\left(\sqrt{1 + \frac{4}{f\left(E\right)}} - 1\right) - 1 .\label{eq:ebic1}
\end{eqnarray}
where the dimensionless factor $f(E)$ is given by:
\begin{eqnarray}
f(E)&=&\left(\frac{L_{\rm drift}\left(E\right)}{L_{\rm diff}}\right)^2
\end{eqnarray}
and the drift length and diffusion length are:
\begin{eqnarray}
L_{\rm drift}\left(E\right) &=& \mu\tau E \\
L_{\rm diff} &=& \sqrt{V_T \mu \tau}
\end{eqnarray}
where $V_T=k_{\rm B}T/q$ is the thermal voltage.  The collection efficiency approaches unity for $f\gg 1$, or when the drift length is much greater than the diffusion length.  The efficiency falls below 1 when $f\approx 1$, or when the drift and diffusion lengths are similar.  Eq. \ref{eq:ebic1} assumes a {\it uniform} electric field.  However we find that using the position-dependent field $E(x)$ of the $p$-$n$ junction to generate a position-dependent efficiency $\eta(E(x))$ results in excellent agreement between the analytical expression and the full numerical simulation.  This is shown in Fig. \ref{fig:model1}(b), demonstrating the applicability of Eq. \ref{eq:ebic1} to situations where recombination in the depletion region is substantial.

Eq. \ref{eq:ebic1} may be inverted to express $\mu\tau$ in terms of the EBIC efficiency $\eta$:
\begin{eqnarray}
\mu\tau=\frac{k_{\rm B}T}{2qE^2}\frac{\left(1+\eta\right)^2}{\left(1-\eta\right)} \label{eq:estimate}
\end{eqnarray}
To recast the above in terms of basic material parameters, we take the magnitude of the maximum field to be $E=V_{bi}/L_W$, where $L_W$ is the depletion width, given by $L_W=\sqrt{2\epsilon V_{bi}/N_A}$.  Here $\epsilon$ is the dielectric constant, $N_A$ is the doping density, and $V_{bi}$ is the built-in potential of the $p$-$n$ junction.  This leads to an estimate for $\mu\tau$ in terms of the maximum efficiency $\eta_{\rm max}$:
\begin{eqnarray}
\mu\tau=\frac{\epsilon k_{\rm B}T}{q^2N_AV_{\rm bi}}\frac{\left(1+\eta_{\rm max}\right)^2}{\left(1-\eta_{\rm max}\right)} \label{eq:estimate}
\end{eqnarray}

We refer the reader to Appendix B for a detailed derivation of Eq. \ref{eq:ebic1}, as well an explicit description of the region within the depletion region for which this expression applies.

\section{comparison to experiment}\label{sec:comparison}

Analyzing the experimental data of Fig. \ref{fig:fit1} with the model of the previous section immediately points to the need for a spatially varying value of $\mu\tau$.  This is seen by using Eq. \ref{eq:estimate} together with the experimental $\eta_{\rm max}=0.5$ to estimate the value of $\mu\tau$ within the depletion region.  For the data presented in Fig. \ref{fig:fit1}, this leads to $\mu\tau \approx 10^{-10}~{\rm cm^2/V}$.  On the other hand, as discussed in Sec. II, the decay length of the EBIC signal in the neutral region is given by the diffusion length $L_D=\sqrt{\left(k_{\rm B}T/q\right)\times\left( \mu \tau\right)}$.  This leads to an estimate of $\mu\tau \approx 10^{-6}~{\rm cm^2/V}$ in the neutral region.  
It is therefore clearly necessary to postulate a spatially varying $\mu\tau$ in order to reproduce the experimental data using the model presented here.

\begin{figure}[h!]
\begin{center}
\vskip 0.2 cm
\includegraphics[width=9.5cm]{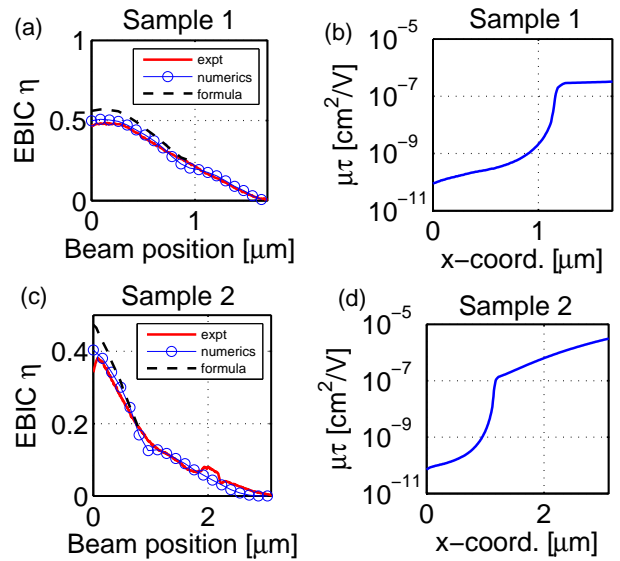}
\vskip 0.2 cm \caption{(a)Comparison between experimental EBIC efficiency for sample 1, and the model result using a profile for $\mu\tau$ as shown in (b), for both numerical simulations and using the formula from Eq. \ref{eq:ebic1}.  (c) shows the same data for Sample 2, with the model using a $\mu\tau$ profile as shown in (d).  The model assumes an electric field corresponding to a doping density of $10^{15}~{\rm cm^{-3}}$ dand a dielectric constant of 11 (see Eq. \ref{eq:ex}).}\label{fig:comparison}
\end{center}
\end{figure}

An estimate of the full spatial dependence of $\mu\tau$ can be made using the experimental data together with Eq. \ref{eq:ebic1}, combined with an assumption of the standard form for the internal field:
\begin{eqnarray}
E\left(x\right) = \frac{N_A}{\epsilon}\left(x-L_W\right) \label{eq:ex}
\end{eqnarray}
Fig. \ref{fig:comparison} shows the spatial variation of $\mu\tau$ which reproduces the experimental data for the two samples, using both Eq. \ref{eq:ebic1} and 1-d numerical simulations.  We find a large variation (over three orders of magnitude) is necessary to achieve quantitative agreement.

We further note that the value of $\mu\tau$ within the depletion region is approximately 2 orders of magnitude smaller than previous estimates.  Ref. \cite{gessert} employs time resolved photoluminescence; using the decay of bandgap and sub-bandgap peaks to estimate the bulk lifetime and drift velocity, respectively, they find $\mu\tau\approx10^{-8}~{\rm cm^2/V}$.
Ref. \cite{tof} utilizes time-of-flight techniques with optical excitation to estimate a similar value of $\mu\tau\approx 10^{-8}~{\rm cm^2/V}$.  We have conducted optical external quantum efficiency measurements, and for short wavelength light where most of the excitation is within the depletion region, we observe an external quantum efficiency of 85\%, leading to a lower limit on $\mu\tau$ of approximately $10^{-9}~{\rm cm^2/V}$ (using Eq. \ref{eq:ebic1}).  Finally, the short circuit current density of the samples 1 and 2 are $23.5~{\rm mA/cm^2}$ and $23.3~{\rm mA/cm^2}$, respectively.  For a bandgap of $1.5~{\rm eV}$ and an incident spectrum from 1 sun illumination, the maximum short circuit current density is approximately $34~{\rm mA/cm^2}$, setting a lower limit on the collection probability of $70\%$.  These considerations lead to the conclusion that the electron beam excitation may be strongly affecting the carrier dynamics, especially in the depletion region.

We next discuss the plausibility and possible origins of the spatially varying and low $\mu\tau$ value implied by the model.  We will argue that the low value may be related to the high generation rate density of the electron beam excitation, and subsequent increased radiative recombination.  We stipulate at the outset that further measurements are required to definitively make such a conclusion.  Nevertheless, at least a portion of the measurements conform semi-quantitatively to this model, as we describe next.

The high generation rate density $G$ of the electron beam combined with a low mobility for CdTe results in a very high nonequilibrium carrier concentration.  Considering $G$ first: for an electron beam current of $200~{\rm pA}$ at energy $5~{\rm keV}$, the total generation rate is $1.5\times10^{12}~{\rm s^{-1}}$ (see Eq. \ref{eq:G}), while the excitation length scale is approximately $120~{\rm nm}$ (see Eq. \ref{eq:GR}).  This leads to a generation rate density of $10^{26}~{\rm 1/cm^3\cdot s}$, or about $10^5$ times greater than the generation rate density under 1 sun illumination.  Additionally, the concentration of nonequilibrium carriers $c_{\rm neq}$ within the excitation volume varies inversely with mobility (slower carriers accumulate more).  For sufficiently large nonequilibrium concentration, radiative and Auger recombination mechanisms become important \cite{fonash}: the lifetime associated with radiative and Auger recombination vary as $1/c_{\rm neq}$ and $1/c_{\rm neq}^2$, respectively.  Here we focus on the radiative component, as our estimates indicate it more likely plays an important role.

A calculation of the full concentration profile requires 3-dimensional modeling, which is beyond the scope of the current paper.  We instead make some analytical estimates assuming that the excitation bulb size is smaller than the diffusion length and the depletion width.  More details of these estimates are provided in Appendix C.  Here we note that for an excitation in the neutral region, the 3-dimensional diffusion equation with a point-source excitation results in a maximum concentration which scales as $G_{\rm tot}/\left(V_T \mu R_B\right)$.  For an excitation in the depletion region, we assume the drift length is much greater than the diffusion length, so that the motion is mostly one-dimensional along the direction of the field, while the transverse area of the excitation (equal to $\pi R_B^2$) is essentially constant.  This results in a maximum concentration which scales as $G_{\rm tot}/\left(\mu E R_B^2\right)$.  The physical picture is that the maximum concentration is larger in the depletion region because the carrier drift precludes the charge from ``spreading out" in the lateral directions (see App. \ref{sec:estimates} for more discussion on these approximations).

We take the mobility in the neutral region $\mu_{\rm neutral}$ to be $50~{\rm cm^2/(V\cdot s})$.  Measured values of the drift mobility are lower, being $15~{\rm cm^2/(V\cdot s)}$ in Ref. \cite{gessert} and $0.7~{\rm cm^2/(V\cdot s)}$ in Ref. \cite{tof}, so we take the mobility in the depletion region to be $\mu_{\rm depletion} = 5~{\rm cm^2/(V\cdot s)}$.  We let $E=10^4~{\rm V/cm}$, and since 75\% of the excitation occurs within 1/3 of the bulb size $R_B$, we use $R_B/3$ for the excitation size.  The radiative coefficient of CdTe has been measured as $B_{\rm radiative}=10^{-9}~{\rm cm^3/ s}$ \cite{Brad}.  This leads to an estimate of $\tau_{\rm neutral} = 1/B_{\rm rad}c^{\rm max}_{\rm neutral} \approx 10^{-8}~{\rm s}$ and $\tau_{\rm depletion} = 1/B_{\rm rad}c^{\rm max}_{\rm depletion} \approx 10^{-10}~{\rm s}$.  The resulting values for $\mu\tau$ are $10^{-10}~{\rm cm^2/V}$ in the depletion region, and $10^{-7}~{\rm cm^2/V}$ in the neutral region.  These values of $\mu\tau$ in turn lead to a maximum efficiency of $\eta_{\rm max}=0.52$ (using Eq. \ref{eq:ebic1}), and a diffusion length in the neutral region $1.1~{\rm \mu m}$, consistent with experiment.  However, we note that the dependence on the beam energy does not conform fully to this picture.  Increasing the beam energy increases the excitation bulb size (see Eq. \ref{eq:GR}), so that the generation rate density decreases by a factor of 100 between our $5~{\rm keV}$ and $20~{\rm keV}$ measurements.  The maximum density should also decrease by a factor between 10 (in the case of predominantly radiative recombination) and 100 (in the case of predominantly defect-mediated recombination).  The lifetime would increase by an identical factor, increasing the efficiency according to Eq. (\ref{eq:ebic1}).   We would expect to see more dramatic increase in efficiency than the modest increase seen experimentally.  On the other hand, preliminary studies of electron beam current dependence show that the efficiency decreases with increasing beam current.  (We find that the efficiency decreases from 0.58 to 0.36 as the beam current is increased from 258 pA to 1290 pA, at a beam energy of 5 keV).

An additional explanation for the lower value of $\mu\tau$ in the depletion region implied by the experiment and the model is inhomogeneity in defect density and strain throughout the thickness of the device.  Ref. \cite{pal} utilizes Auger electron spectroscopy for depth profiling of CdS-CdTe photovoltaics, and observes substantial S interdiffusion in the depletion region, although S preferentially diffuses along grain boundaries \cite{herndon}.  Ref. \cite{halliday} utilizes Secondary Ion Mass Spectroscopy to observe an inhomogeneous distribution of impurities throughout the device thickness, as a function of ${\rm CdCl_2}$ activation.  This impurity distribution should lead to a position-dependent carrier lifetime from Shockley-Read-Hall recombination.

We reiterate that these hypotheses are necessitated by the two very basic features of the data: the reduced maximum EBIC efficiency (requiring a low $\mu\tau$ in the depletion region), and a nonzero EBIC efficiency that extends into the neutral region (requiring a higher $\mu\tau$ there).  In our experience, these features of the data are unique to CdTe.  As mentioned in Sec. \ref{sec:expt}, experiments on Si show a maximum collection efficiency of 1.  We have also performed EBIC experiments on CIGS samples, and also find a maximum collection efficiency of 0.85, which is quite close to 1 within the experimental uncertainty (for electron beam current of $59~{\rm pA}$ and energy $5~{\rm eV}$).  A key difference between CdTe and these other materials may result from the unique processing conditions of CdS-CdTe photovoltaics.  In particular, the annealing step of device preparation takes place after CdS deposition in CdTe (in contrast to CIGS).  This leads to nonuniform grain boundary passivation and increased stress near the CdS-CdTe interface.  A reduced mobility in the junction will result in a larger maximum concentration, so that other recombination mechanisms become important, as described previously.


More work is needed to determine definitively the source of the reduced EBIC efficiency.  The influence of radiative and Auger recombination can be minimized by reducing the generation rate density $G$.  $G$ may be reduced by lowering the electron beam current or by increasing the electron beam energy (so that the excitation bulb volume increases).  Both approaches pose challenges however: increasing the beam energy leads to excitation bulbs which enclose multiple grains/grain boundaries.  A quantitative account of the EBIC response of grain boundaries requires detailed modeling, so that the interpretation of high beam energy EBIC is not straightforward (see discussion in Sec. \ref{sec:expt}).  Lowering the beam current for small beam energies is limited by signal to noise considerations.  An opposite approach is to systematically {\it increase} the beam current in order to quantify the relative contributions of radiative and Auger recombination.  This is done in Ref. \cite{durose}, which considers the effect of high level injection explicitly in the interpretation of grain boundary contrast, and observes a substantial reduction in efficiency with increasing beam current, consistent with the picture presented here.  On the other hand, a non-uniform $\mu\tau$ due to a spatial distribution of defects may be determined by quantitative modeling and analysis of optical EQE data.  In this case it's necessary to measure (or model) the optics realistically in order to determine the internal quantum efficiency, and to perform modeling to deduce the collection probability function $\phi(x)$.

\section{Conclusion}

We've presented a critical examination of cross-sectional EBIC data on CdS-CdTe photovoltaics, with the main experimental observation that the maximum collection efficiency is less than $1$.  This indicates either very low values for $\mu\tau$, or screening of the built-in field.  This work focuses on the former scenario, with a theoretical result of an expression for the EBIC efficiency when recombination in the depletion region can't be ignored.  This model is consistent with experimental results only if there is spatial variation in $\mu\tau$ throughout the depletion region.  Application of this model leads to values of $\mu\tau$ which are quite low (on the order of $10^{-10}~{\rm cm^2/V}$).  We speculate that the high generation rate density associated with the electron beam excitation may drive the system to a regime in which radiative recombination processes become important, leading to a reduced value of the carrier lifetime.  We consider this work as a step towards a fuller and more quantitative understanding of the EBIC response of polycrystalline materials.  The maximum collection efficiency as a key parameter which must be understood in order to have confidence in any comprehensive model of the system.  Future work will include a more expanded study of possible high injection effects, as well as more realistic 3-dimensional modeling efforts.

\section*{Acknowledgment}
H. Yoon acknowledges support under the Cooperative Research Agreement between the University of Maryland and the National Institute of Standards and Technology Center for Nanoscale Science and Technology, Award 70NANB10H193, through the University of Maryland.  P. K. and R. W. C. were supported by the DOE/NSF F-PACE Program (Contract DE-EE0005405).

\begin{appendix}
\section{Experimental details}

\begin{figure}[h!]
\begin{center}
\vskip 0.2 cm
\includegraphics[width=7cm]{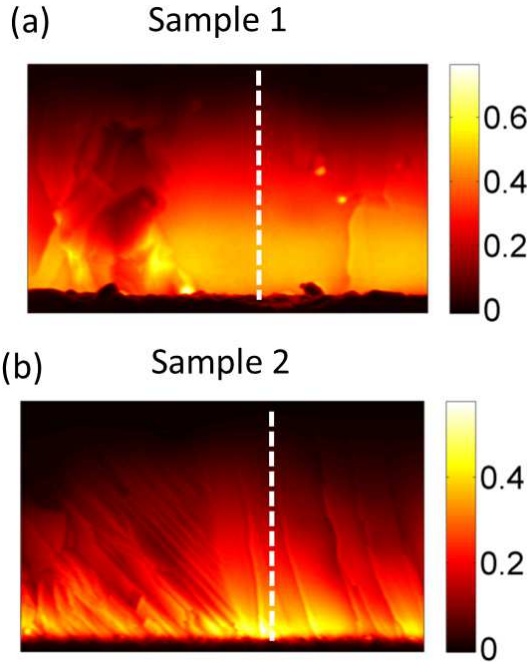}
\vskip 0.2 cm \caption{EBIC efficiency maps of samples 1 and 2.  The white dashed lines correspond to the traces used for the fitting described in the paper.  The traces are chosen to minimize the effect of grain boundaries on the signal.  The field of view is $2.1~{\rm \mu m}\times3.1~{\rm \mu m}$ in (a) and $3.7~{\rm \mu m}\times6~{\rm \mu m}$  in (b). }\label{fig:maps}
\end{center}
\end{figure}

Fig. \ref{fig:maps} shows the full spatially resolved EBIC response at a beam energy of $5~{\rm keV}$ for the two samples.  The structure of the signal is due to surface roughness and grain boundaries, although it is difficult to determine which is predominant {\it a priori}.
The treatment of surface effects is the topic of future work.  The white dashed lines indicate the scan which is used in the data of Fig. \ref{fig:fit1} of the main text.

\section{Mathematical model}

\begin{figure}[h!]
\begin{center}
\vskip 0.2 cm
\includegraphics[width=8.5cm]{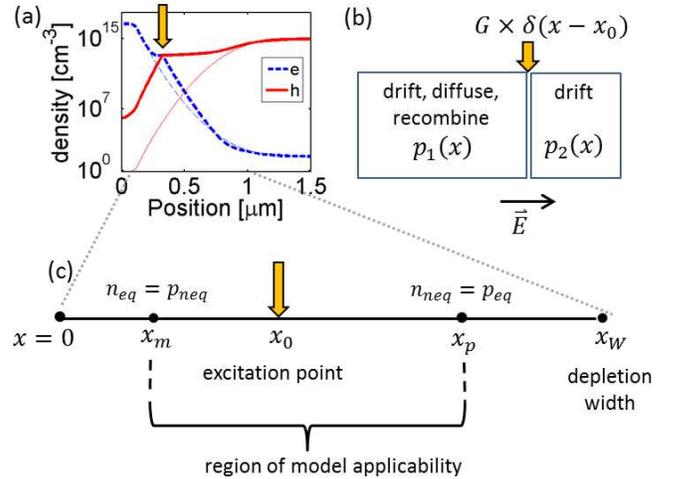}
\vskip 0.2 cm \caption{Model results for a point-source excitation at x=0.35 ${\rm \mu m}$.  (a) shows the density of electrons (blue dashed) and holes (solid red). The thinner lines are the equilibrium densities, while the thicker lines are the total densities upon excitation.  The model parameters are: doping density of p-type material is $N_A=10^{15}~{\rm cm^{-3}},$ majority and minority carrier mobilities are $~\mu=10~{\rm cm^2/(V\cdot s)}$, bulk lifetime is $\tau_{\rm bulk}=50~{\rm ns},$ bandgap is $E_g=1~{\rm eV},$ and dielectric constant $\epsilon=11~\epsilon_0$, where $\epsilon_0$ is the permettivity of free space.  The junction is located at $x=0.12~{\rm \mu m}$, and the total system length is $3~{\rm \mu m}$.  (b) shows a schematic of the analytic model which is intended to capture the important physics of the system.  This model applies only within the depletion region, where the electron and hole are approximately equal at the excitation point.}\label{fig:cartoon}
\end{center}
\end{figure}

We present the derivation of Eq. \ref{eq:ebic1} from the main text.  Fig. \ref{fig:cartoon}(a) shows the result of numerical simulation of a $p$-$n$ junction with a delta-function excitation at $x=0.35~{\rm \mu m}$.  We use this simulation to inform the assumptions we make for the analytical model.  Our first assumption is that the density of electrons and holes are equal at the excitation point.  The region for which this is satisfied is given at the end of this section.  The electron beam induced excitation is modeled as a delta-function at position $x=x_0$.  Here we present an analysis of the induced hole carriers - the treatment of electrons is identical.  For the geometry of Fig. \ref{fig:cartoon}, holes are minority carriers to the left of the excitation ($x<x_0$), and undergo drift, diffusion, and recombination.  To the right of the excitation ($x>x_0$), holes are majority carriers and simply undergo drift.  The schematic of the resulting model for holes is shown in Fig. \ref{fig:cartoon}(b).  Solving the drift-diffusion equation for electron/hole density permits the calculation of the EBIC efficiency.

Here we just consider the behavior of the holes - the treatment of electrons is identical.  The equation of continuity holes to the left of the excitation is:
\begin{eqnarray}
\partial_x \left(\mu p_1 E - D \partial_x p_1\right) = -\frac{p}{\tau_{\rm eff}}
\end{eqnarray}
where $D$ is the hole diffusivity, and $\tau_{\rm eff}=\beta\tau_{\rm bulk}$. $\beta$ varies between 1 (if $p \ll n$) and 2 (if $p=n$); this follows from the form of Read-Shockley-Hall recombination.  We take $\beta=1.75$ for all calculations; this is appropriate to describe the recombination in regions for which the electron and hole density is similar, though not identical.  We find that with this choice, the analytical model reproduces the numerical simulations well.  The continuity equation for holes in the region to the right of the excitation is:
\begin{eqnarray}
\partial_x J_2 = \partial_x \left(\mu p_2 E\right) = 0 \label{eq:j2}
\end{eqnarray}
We assume that the electric field varies slowly compared to the variation of the charge densities, so that $\partial_x E$ is negligible.  The solution is specified by three boundary conditions:  1. the carrier density goes to 0 as $x\rightarrow -\infty$,  2. the density is continuous at the excitation point  $x=x_0$, and 3. the current is discontinuous at the excitation point:
\begin{eqnarray}
p_1(x_0) &=& p_2(x_0)\\
J_1(x_0) - J_2(x_0) &=& G_{\rm tot}
\end{eqnarray}
The solution $p_1(x)$ (where $x<x_0$) which satisfies these boundary conditions is:
\begin{eqnarray}
p_1(x) = \left(\frac{2G_{\rm tot}}{\mu E\left(x\right)}\right)\frac{1}{1+\sqrt{1+4/f\left(x\right)}}~~~~~~~~~~~~~~~\nonumber\\ ~ \times\exp\left[\frac{-\left(x_0-x\right)q E\left(x\right)}{2k_{\rm B}T}\left(1+\sqrt{1+\frac{4}{f\left(x\right)}}\right) \right] \label{eq:px}
\end{eqnarray}
where
\begin{eqnarray}
f\left(x\right)=\frac{q\mu \tau_{\rm eff} E^2\left(x\right)}{k_{\rm B}T} = \left(\frac{L_{\rm drift}(x)}{L_{\rm diff}}\right)^2
\end{eqnarray}
$f(x)$ is the dimensionless parameter which determines the relevance of junction recombination.  For high efficiency solar cells, $f\gg 1$ and junction recombination is negligible.  The solution $p_2(x)$ is spatially constant: $p_2(x)=p_1(x_0)$.

The total recombination $R_{\rm tot}$ is readily determined from the minority carrier density:
\begin{eqnarray}
R_{\rm tot} = 2\int_{-\infty}^{x_0} \frac{p_1(x)}{\tau_{\rm eff}} dx \label{eq:rtot}
\end{eqnarray}
Note that we neglect the spatial dependence of $E(x)$ in performing the integral, letting $E(x)\rightarrow E(x_0)$.  This is justified because the integrand is dominated by contributions near $x=x_0$.  The factor of 2 arises because an equivalent treatment of electrons applies, doubling the recombination contribution from the holes presented here.  The EBIC efficiency is given by $\left(G_{\rm tot}-R_{\rm tot}\right)/G_{\rm tot}$.  This leads to the following form of the EBIC response:
\begin{eqnarray}
\eta\left(x_0\right)=1-\frac{R_{\rm tot}}{G_{\rm tot}} =  f\left(x_0\right)\left(\sqrt{1 + \frac{4}{f\left(x_0\right)}} - 1\right) - 1 \label{eq:ebic}
\end{eqnarray}
Eq. \ref{eq:ebic} is the main theoretical result of the paper.  Notice the transition to an unphysical $\eta<0$ for $f<1/2$.  The reason for this is that for small electric fields (small $f$), the diffusion current in the majority region is comparable to the drift current, violating an assumption of the model (Eq. (\ref{eq:j2})).   In short, the model is not designed to describe small electric fields.

\begin{figure}[h!]
\begin{center}
\vskip 0.2 cm
\includegraphics[width=7cm]{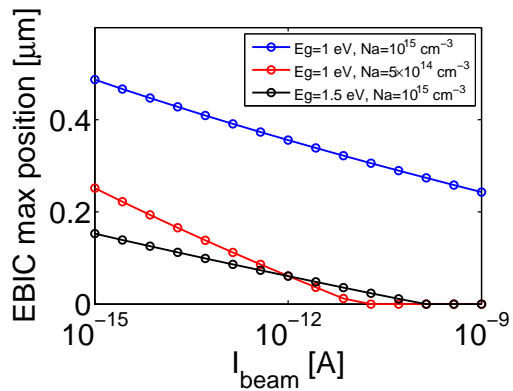}
\vskip 0.2 cm \caption{Location of EBIC maximum measured from the metallurgical junction versus beam current for different material parameters.}\label{fig:maxpos}
\end{center}
\end{figure}

We reiterate Eq. \ref{eq:ebic} applies for excitation rates for which nonequilibrium densities exceed the local majority carrier density.  We denote the minimum value of $x$ for which this holds by $x_m$ (see Fig. \ref{fig:cartoon}(c)).  This is the point with the largest electric field for which the model applies, and corresponds to the position of the maximum EBIC efficiency.  To specify $x_m$, we first write the general expression for the equilibrium electron density $n_{\rm eq}$:
\begin{eqnarray}
n_{\rm eq}(x) &=& n_0\exp\left(\frac{\left(x-L_W\right)^2}{x_d^2}\right) \label{eq:nx}
\end{eqnarray}
where $L_W =\sqrt{2\epsilon V_{\rm bi}/q N_A}$ is the depletion width, $x_d=\sqrt{2\epsilon k_{\rm B}T/q^2 N_A}$ is the Debye length, and $n_0=N_c N_v/N_A \exp\left(-E_g/k_{\rm B}T\right)$.  The nonequilibrium hole density $p_{\rm neq}$ is maximized at the excitation point $x_0$.  Letting $p\left(x_0\right)=n_{\rm eq}(x)$ determines $x_m$:
\begin{eqnarray}
n_0\exp\left(\frac{\left(x_m-L_W\right)^2}{x_d^2}\right)=~~~~~~~~~~~~~~~~~~~~~~~~~~~~~~~ \nonumber \\
 \frac{2G_{\rm tot}}{\left(R\right)^{dim-1}\mu E\left(x_m\right)\left(1+\sqrt{1+4/f(x_m)}\right)} \label{eq:xm}
\end{eqnarray}
where ${dim}$ is dimensionality of the system.  In the present analysis, ${dim=1}$.  Experimental comparisons require $dim=3$.  Eq. \ref{eq:xm} does not admit a closed form solution for $x_m$.  We plot $x_m$ for three different system parameterizations in Fig. \ref{fig:maxpos}.  As the beam current is reduced, the region for which the nonequilibrium density exceeds the equilibrium majority density shrinks, and the EBIC maximum position approaches the neutral point, where $n_{\rm eq}(x)=p_{\rm eq}(x)$.  As the beam current is increased, the nonequilibrium density increases and the maximum position is shifted towards regions of higher field, nearer $x=0$.


For excitation position $x_0<x_m$, the EBIC efficiency decays exponentially with a length scale $\mu\tau E(x_0)$
\begin{eqnarray}
\eta_2(x_0) = \eta \left(x_m\right)\exp\left(-\mu \tau \left(x_m-x_0\right) E(x)\right)  \label{eq:eta2}
\end{eqnarray}
Similar considerations apply for specifiying $x_p$, or the maximum position for which Eq. \ref{eq:ebic} applies.

\section{Estimates of lifetime}\label{sec:estimates}
To determine the lifetimes associated with radiative recombination, the absolute value of the concentration is required.  Here we describe in more detail the analytical estimates used to approximate the maximum carrier concentration and resulting lifetimes.  For excitations in the neutral region, the 3-dimensional diffusion equation with a delta-function excitation at the origin leads is:
\begin{eqnarray}
\frac{1}{r^2}\frac{\partial}{\partial r}\left(r^2\frac{\partial c}{\partial r}\right) = \frac{c(r)}{L_D^2} + \left(\frac{G_{\rm tot}}{D}\times\delta\left(r\right)\right)
\end{eqnarray}
Notice here we assume the recombination is first-order in minority carrier density.  Strictly speaking, this is inconsistent with the conclusion that radiative recombination is dominant.  However, this assumption simplifies the math considerably, and enables an estimate of the threshold generation rate for which radiative recombination becomes important.  The delta-function imposes a boundary condition on c(4):
\begin{eqnarray}
\lim_{r\rightarrow 0} -4\pi r^2 D\frac{\partial c}{\partial r} = G_{\rm tot} \label{eq:bcsphere}
\end{eqnarray}
The solution is given by:
\begin{eqnarray}
c(r)=\frac{G_{\rm tot}}{4\pi D r}\times\exp\left(-r/L_D\right)
\end{eqnarray}
$c(r)$ diverges as $r\rightarrow 0$, so a physical cutoff of the excitation bulb size $R_B$ is used to calculate the maximum density of nonequilibrium carrier density.  We also suppose that $R_B \ll L_D$, leading to:
\begin{eqnarray}
c^{\rm max}_{\rm neutral}\approx \frac{G_{\rm tot}}{4\pi V_T \mu_{\rm neutral} R_B}.
\end{eqnarray}
Notice that the maximum concentration is {\it independent} of $L_D$.  For $R_B \ll L_D$, the maximum concentration is set by the structure of the divergence in 3-dimensions: according to Eq. \ref{eq:bcsphere}, the diffusion current immediately outside the localized excitation spot must carry away carriers at the same rate as their generation.  This requires a sharp gradient in concentration, and associated large value of concentration very near the excitation point.

For completeness, we also give the general expression for the maximum density, valid for any $R_B/L_D$:
\begin{eqnarray}
c_{\rm neutral}^{\rm max}=G\tau \left(1-\left(\frac{L_D+R_B}{L_D}\right)\exp\left(-R_B/L_D\right)\right)
\end{eqnarray}
For large excitations $R_B\gg L_D$, the above expression yields $c_{\rm neutral}^{\rm max}\approx G\tau$, as expected.  As discussed in Sec. \ref{sec:comparison}, one regime of high injection is entered when the effective radiative recombination lifetime $B_{\rm rad}/c^{\rm max}$ is smaller than the defect-related recombination lifetime.  (Other regimes of high injection correspond to total generation rates which exceed the current density afforded by the material doping and built-in voltage.  This will be detailed in forthcoming work, and is explored in Ref. \cite{Nichterwitz}.)

For excitations within the depletion region, we assume that the drift time is less than the diffusion time, so that the motion of the nonequilibrium charge is essentially 1-dimensional along the direction of the field, while the transverse area of the excitation (equal to $\pi R^2$) is essentially constant.  This is equivalent to assuming $f\gg1$.  By fitting the data, we find $f\approx 3$, so that this approximation is of moderate validity.  Solving the 1-dimensional drift equation for constant field $E$ leads to a concentration given by $c=G_{\rm tot}/\left(\mu E\right)$.  In 3-dimensions, the maximum concentration is then given by:
\begin{eqnarray}
c^{\rm max}_{\rm depletion}=\frac{G_{\rm tot}}{\pi\mu_{\rm depletion} R_B^2 E}
\end{eqnarray}
Given these expressions for the maximum concentration, the lifetime due to radiative recombination is readily determined by: $\tau=1/\left(B_{\rm rad}c^{\rm max}\right)$.  Plugging in the estimates for material parameters as given in Sec. \ref{sec:comparison} leads to the values of $\mu\tau$ given in the main text.


\end{appendix}


\begin{thebibliography}{1}
\bibitem{footnote}
Electrons may undergo a series of inelastic scattering events - generating electron-hole pairs along the way - before being re-emitted from the surface.  It's therefore necessary to use the reflected energy (not reflected number) as the backscattering coefficient in Eq. \ref{eq:G}.
\bibitem{phi_footnote} We use the same form of $\phi\left(x\right)$ as Ref. \cite{rau}, except with an extra parameter $\phi_0$, as described in the text.  For excitations in the neutral region, $\phi(x)$ is given as (see main text for description of parameters):
\begin{eqnarray}
\frac{\phi(x)}{\phi_0}=\frac{1/L_{D}'\cosh\left(\frac{x-L}{L_D'}\right)-S_C\sinh\left(\frac{x-L}{L_D'}\right)}
{1/L_{D}'\cosh\left(\frac{L_W-L}{L_D'}\right)-S_C\sinh\left(\frac{L_W-L}{L_D'}\right)}
\end{eqnarray}
\end{thebibliography}
\end{document}